
\message
{JNL.TEX version 0.92 as of 6/9/87.  Report bugs and problems to Doug Eardley.}

\catcode`@=11
\expandafter\ifx\csname inp@t\endcsname\relax\let\inp@t=\input
\def\input#1 {\expandafter\ifx\csname #1IsLoaded\endcsname\relax
\inp@t#1%
\expandafter\def\csname #1IsLoaded\endcsname{(#1 was previously loaded)}
\else\message{\csname #1IsLoaded\endcsname}\fi}\fi
\catcode`@=12







\def\beginlinemode{\endmode
  \begingroup\parskip=0pt \obeylines\def\\{\par}\def\endmode{\par\endgroup}}
\def\beginparmode{\endmode
  \begingroup \def\endmode{\par\endgroup}}
\let\endmode=\par
{\obeylines\gdef\
{}}
\def\singlespace{\baselineskip=\normalbaselineskip}

\def\oneandahalfspace{\baselineskip=\normalbaselineskip
  \multiply\baselineskip by 3 \divide\baselineskip by 2}
\def\doublespace{\baselineskip=\normalbaselineskip \multiply\baselineskip by 2}






\def\title			
  {\null\vskip 3pt plus 0.2fill
   \beginlinemode \doublespace \raggedcenter \bf}

\def\author			
  {\vskip 3pt plus 0.2fill \beginlinemode
   \singlespace \raggedcenter\sc}

\def\affil			
  {\vskip 3pt plus 0.1fill \beginlinemode
   \oneandahalfspace \raggedcenter \sl}

\def\abstract			
  {\vskip 3pt plus 0.3fill \beginparmode
   \oneandahalfspace ABSTRACT: }

\def\endtitlepage		
  {\endpage			
   \body}

\def\body			
  {\beginparmode}		

\def\beginitems{
\par\medskip\bgroup\def\i##1 {\item{##1}}\def\ii##1 {\itemitem{##1}}
\leftskip=36pt\parskip=0pt}
\def\enditems{\par\egroup}

\def\beneathrel#1\under#2{\mathrel{\mathop{#2}\limits_{#1}}}

\def\refto#1{~[{#1}]}

\def\references			
  {
   \beginparmode
   \frenchspacing \parindent=0pt \leftskip=1truecm
   \parskip=8pt plus 3pt \everypar{\hangindent=\parindent}}

\gdef\refis#1{\item{#1.\ }}			

\gdef\journal#1, #2, #3, 1#4#5#6{		
    {\sl #1~}{\bf #2}, #3 (1#4#5#6)}		

\def\endreferences{\body}

\catcode`@=11
\newcount\r@fcount \r@fcount=0
\newcount\r@fcurr
\immediate\newwrite\reffile
\newif\ifr@ffile\r@ffilefalse
\def\w@rnwrite#1{\ifr@ffile\immediate\write\reffile{#1}\fi\message{#1}}

\def\writer@f#1>>{}
\def\referencefile{
  \r@ffiletrue\immediate\openout\reffile=\jobname.ref%
  \def\writer@f##1>>{\ifr@ffile\immediate\write\reffile%
    {\noexpand\refis{##1} = \csname r@fnum##1\endcsname = %
     \expandafter\expandafter\expandafter\strip@t\expandafter%
     \meaning\csname r@ftext\csname r@fnum##1\endcsname\endcsname}\fi}%
  \def\strip@t##1>>{}}

\def\citeall#1{\xdef#1##1{#1{\noexpand\cite{##1}}}}
\def\cite#1{\each@rg\citer@nge{#1}}	

\def\each@rg#1#2{{\let\thecsname=#1\expandafter\first@rg#2,\end,}}
\def\first@rg#1,{\thecsname{#1}\apply@rg}	
\def\apply@rg#1,{\ifx\end#1\let\next=\relax
\else,\thecsname{#1}\let\next=\apply@rg\fi\next}

\def\citer@nge#1{\citedor@nge#1-\end-}	
\def\citer@ngeat#1\end-{#1}
\def\citedor@nge#1-#2-{\ifx\end#2\r@featspace#1 
  \else\citel@@p{#1}{#2}\citer@ngeat\fi}	
\def\citel@@p#1#2{\ifnum#1>#2{\errmessage{Reference range #1-#2\space is bad.}%
    \errhelp{If you cite a series of references by the notation M-N, then M and
    N must be integers, and N must be greater than or equal to M.}}\else%
 {\count0=#1\count1=#2\advance\count1 by1\relax\expandafter\r@fcite\the\count0,%
  \loop\advance\count0 by1\relax
    \ifnum\count0<\count1,\expandafter\r@fcite\the\count0,%
  \repeat}\fi}

\def\r@featspace#1#2 {\r@fcite#1#2,}	
\def\r@fcite#1,{\ifuncit@d{#1}
    \newr@f{#1}%
    \expandafter\gdef\csname r@ftext\number\r@fcount\endcsname%
                     {\message{Reference #1 to be supplied.}%
                      \writer@f#1>>#1 to be supplied.\par}%
 \fi%
 \csname r@fnum#1\endcsname}
\def\ifuncit@d#1{\expandafter\ifx\csname r@fnum#1\endcsname\relax}%
\def\newr@f#1{\global\advance\r@fcount by1%
    \expandafter\xdef\csname r@fnum#1\endcsname{\number\r@fcount}}

\let\r@fis=\refis			
\def\refis#1#2#3\par{\ifuncit@d{#1}
   \newr@f{#1}%
   \w@rnwrite{Reference #1=\number\r@fcount\space is not cited up to now.}\fi%
  \expandafter\gdef\csname r@ftext\csname r@fnum#1\endcsname\endcsname%
  {\writer@f#1>>#2#3\par}}

\def\ignoreuncited{
   \def\refis##1##2##3\par{\ifuncit@d{##1}%
     \else\expandafter\gdef\csname r@ftext\csname r@fnum##1\endcsname\endcsname%
     {\writer@f##1>>##2##3\par}\fi}}

\def\r@ferr{\endreferences\errmessage{I was expecting to see
\noexpand\endreferences before now;  I have inserted it here.}}
\let\r@ferences=\references
\def\references{\r@ferences\def\endmode{\r@ferr\par\endgroup}}

\let\endr@ferences=\endreferences
\def\endreferences{\r@fcurr=0
  {\loop\ifnum\r@fcurr<\r@fcount
    \advance\r@fcurr by 1\relax\expandafter\r@fis\expandafter{\number\r@fcurr}%
    \csname r@ftext\number\r@fcurr\endcsname%
  \repeat}\gdef\r@ferr{}\endr@ferences}


\let\r@fend=\endpaper\gdef\endpaper{\ifr@ffile
\immediate\write16{Cross References written on []\jobname.REF.}\fi\r@fend}

\catcode`@=12

\citeall\refto		

\input epsf.tex
\magnification 1200
\vsize=7.8in
\hsize=5.7in
\voffset=0.1in
\hoffset=-0.15in
\newcount\eqnumber
\baselineskip 18pt plus 0pt minus 0pt


\font\rmmthree=cmbx10 scaled 1500
\font\rmmtwo=cmbx10 scaled 1200
\font\rmmoneB=cmbx10 scaled 1100

\font\rmmoneI=cmti10 scaled 1000

\font\ninerm=cmr10 scaled 900

\font\eightit=cmti10 scaled 800
\font\eightrm=cmr10 scaled 800
\font\eightbf=cmbx10 scaled 800


\def\title#1{\centerline{\noindent{\rmmthree #1}}\nobreak\smallskip\eqnumber=1}

\def\sectbegin#1#2{\bigskip\bigbreak\leftline{\rmmtwo
#1~#2}\nobreak\medskip\nobreak}

\def\nosectbegin#1{\bigskip\bigbreak\leftline{\rmmtwo #1}\nobreak\medskip}


\def\lapp{\hbox{$ {     \lower.40ex\hbox{$<$}
                   \atop \raise.20ex\hbox{$\sim$}
                   }     $}  }
\def\gapp{\hbox{$ {     \lower.40ex\hbox{$>$}
                   \atop \raise.20ex\hbox{$\sim$}
                   }     $}  }

\def\marbul{\strut\vadjust{\kern-2pt$\bullet$}}

\def\rr{\rangle}
\def\ll{\langle}

\def\specialwarn{\vtop to
\strutdepth{\baselineskip\strutdepth\vss\llap{
\lower.1ex\hbox{$\bigtriangleup$}\kern-0.884em$\triangle$\kern-0.5667em{\eightrm
!}\hskip 13.5pt}\null}}
\def\strutdepth{\dp\strutbox}


\def\new{{\the\eqnumber}\global\advance\eqnumber by 1}
\def\delaynew{{\the\eqnumber}}
\def\nownew{\global\advance\eqnumber by 1}
\def\last{\advance\eqnumber by -1 {\the\eqnumber}
    \global\advance\eqnumber by 1}
\def\eqnam#1{
\xdef#1{\the\eqnumber}}


\def\figure#1#2#3#4#5#6{
\topinsert
\null\hskip #4\relax
\epsfxsize #2
\epsfysize #3
\epsfbox{#1}
\medskip
{\baselineskip 10pt\noindent\narrower\rm\hbox{\eightbf
Figure #5}:\quad\eightrm
#6 \smallskip}
\endinsert
}


\def\dalemb#1#2{{\vbox{\hrule height .#2pt
\hbox{\vrule width.#2pt height#1pt \kern#1pt\vrule width.#2pt}
\hrule height.#2pt}}}

\def\tdot{\kern -8.5pt {}^{{}^{\hbox{...}}}}
\def\dotprime{\kern -8.0pt{}^{{}^{\hbox{.}~\prime}}}

\def\foot#1#2{\footnote{#1}{\eightrm #2 \vskip -10pt}}


\title{SMALL-ANGLE ANISOTROPIES IN THE CMBR}
\title{FROM ACTIVE SOURCES}
\vskip 15pt
\centerline{\rmmoneB R.$\,$A. Battye}
\vskip 15pt
\centerline{\rmmoneI Theoretical Physics Group, Blackett Laboratory, Imperial College}
\centerline{\rmmoneI Prince Consort Road, London SW7 2BZ, U.K.
\foot{}{\smallskip\noindent Email : r.battye@ic.ac.uk \smallskip
\noindent Paper submitted to {\eightit Physical Review D}.}}
\vskip 10pt
\medskip \centerline{\rmmoneB Abstract} 
\medskip 
{\narrower{\baselineskip 9pt \ninerm \noindent 
We consider the effects of photon diffusion on the small-angle microwave background anisotro\-pies due to active source models. We find that fluctuations created just before the time of last scattering allow anisotropy to be created on scales much smaller than allowed by standard Silk damping. Using simple models for string and texture structure functions as examples, we illustrate differences in the angular power spectrum  of the intrinsic and Doppler components of the anisotropy on  scales of order a few arcminutes. In particular, we find that the Doppler peak heights are modified by 10-50\% and small-angle fall-off is power law rather than exponential.}\smallskip}   
\bigskip

\sectbegin{1.}{Introduction}

\noindent Accurate measurements of the anisotropy in the cosmic microwave background radiation are a powerful probe of modern cosmological models since they represent a snapshot of the universe just before it became transparent to electromagnetic radiation about 400,000 years after the Big Bang. The standard picture is that the initial seed fluctuations were created close to the Planck epoch by quantum effects, during a period of super-luminal expansion known as inflation. These perturbations have been termed as `passive'\refto{ACFM} since they were created during an early epoch and follow a linear deterministic evolution until last scattering, where they leave an imprint in the microwave background.

There are, however, a potentially much larger group of theories, known as `active' source models\refto{MACFa,MACFb}, which could lead to the creation of microwave anisotropies and  the formation of large-scale structure. The best motivated of these are those associated with topological defects\refto{kibb,VS,HK} formed during cosmological phase transitions, such as cosmic strings\refto{zeld,vil} and textures\refto{turok}; their generic feature being that they create perturbations over a very wide range of scales at all times, from their formation to the present day.

The large-angle $(\gapp1^{\circ})$ anisotropies due to defects have been shown to be compatible with those detected by COBE since they lead to a near scale-invariant spectrum of perturbations on the relevant scales (see, for example, \refto{ACSSV}), while little quantitative work exists on the formation of large-scale structure due to calculational difficulties in modelling non-linear effects. The small-angle $(\lapp 1^{\circ})$ microwave background anisotropies provide an area where the predictions from passive and active theories can be tested since they are generically very different. A number of works\refto{ACFM,MACFa,MACFb,DGS,CT} have predicted shifts in the positions of the so called Doppler or Sakharov peaks which occur largely due to the differences in the tight coupling solutions. Here, we shall discuss the effects of photon diffusion or Silk damping\refto{silk} which have until now been  ignored or introduced {\it ad hoc}. 

The crucial period for understanding these effects is that just before last scattering. During this epoch the acoustic oscillations in the photon-baryon fluid are damped by the increasing mean free path of the photons. It is not difficult to see that fluctuations created  after the onset of this regime will receive less damping than those created before it. In particular, those created just before the time of last scattering will receive virtually no damping at all. If perturbations are created on all scales above the defect size, as is thought to be the case for defect models, then it will be possible for anisotropy on small angular scales to remain to the present day. We will see that it is not sufficient to model these effects with a simple multiplication by an exponential suppression factor across all scales. Rather it requires careful consideration of the time at which fluctuations are created. Simple estimates will show that there are potentially important modifications to peak heights and also power-law suppression at small angular scales, rather than exponential.

\sectbegin{2}{Analytic formalism}

\noindent We shall use the analytic formalism developed in refs.\refto{HSa,HSb} to describe the small-angle microwave background anisotropies for passive perturbations. There are, however, shortcomings in the way Silk damping is modelled which will lead to discrepancies at small angular scales when applied to active sources. The starting point for this treatment is the first-order collisional Boltzmann equation in Fourier space ignoring the effects of polarisation,
\eqnam{\boltzmann}
$$\dot\Theta+ik\mu(\Theta+\Psi)=-\dot\Phi+\dot\kappa\left(\Theta_0-\Theta-\textstyle{1\over 10}P_2(\mu)\Theta_2-iP_1(\mu)V_{\rm b}\right)\,.\eqno(\new)$$
In this equation, $\Theta(k,\eta,\mu)=\sum (-i)^l\Theta_l(k,\eta)P_l(\mu)$ is the Newtonian temperature perturbation, $\eta$ is conformal time, $k$ is the wavenumber, $P_l(\mu)$ is the Legendre polynomial associated with angular variable $\mu=\cos\theta$, $V_{\rm b}$ is the baryon velocity, $\dot\kappa$ is the differential optical depth due to Thomson scattering and $\Psi\,,\,\Phi$ are the gauge invariant potentials\refto{bar} assumed to be characterised by an external source. One can solve this equation for $\Theta_l$ $(l>1)$, if one allows the intrinsic (or monopole) temperature anisotropy and the baryon velocity to act as extra source terms. The solution at the present day is then\refto{HSc}
\eqnam{\freestream}
$$\eqalign{&{\Theta_l(k,\eta_0)\over 2l+1}=\int_0^{\eta_0}d\eta \,e^{-\kappa(\eta,\eta_0)}\left(
\dot\Psi-\dot\Phi\right)j_l[k\Delta\eta]+\int_0^{\eta_0}d\eta\,
\dot\kappa e^{-\kappa(\eta,\eta_0)}\left(\Psi-\Phi\right)j_l[k\Delta\eta]
\cr &+\int_0^{\eta_0}d\eta\,\dot\kappa e^{-\kappa(\eta,\eta_0)}\left(
\widehat\Theta_0 j_l[(k\Delta\eta]+{l\over 2l+1}
V_{\rm b}j_{l-1}[k\Delta\eta]-{l+1\over 2l+1}V_{\rm b}
j_{l+1}[k\Delta\eta]\right)\,,}\eqno(\new)$$
where $\widehat\Theta_0=\Theta_0+\Phi$, $\Delta\eta=\eta_0-\eta$, $\kappa(\eta_1,\eta_2)=\int_{\eta_1}^{\eta_2}d\eta\dot\kappa(\eta)$ and $j_l(x)$ is a spherical Bessel function. This expression is slightly  different to previous formulations, in that there are the separate terms $\Theta_0+\Phi$ and $\Psi-\Phi$, as opposed to just $\Theta_0+\Psi$. The reason for this is that, as we will see, it is only $\Theta_0+\Phi$, henceforth described as the intrinsic anisotropy, which is modified by Silk damping. 

The function $\dot\kappa e^{-\kappa(\eta,\eta_0)}$ is known as the `visibility function' and is sharply peaked around the time of last scattering $\eta_*$, while $e^{-\kappa(\eta,\eta_0)}$ is zero for $\eta<<\eta_*$ and effectively constant $\eta>>\eta_*$.
Therefore, one can see that the anisotropy is essentially due to three effects. The first sums up all contributions due to the motion of the sources after the time of last scattering and the second is just the difference of the two potentials at last scattering. The combination of these two terms, 
known as the  Sachs-Wolfe effect\refto{SW}, dominates the anisotropy on large angular scales and possibly again at very small $(\lapp 1^{\prime})$ angular scales where, for example, the Kaiser-Stebbins effect\refto{KS} is prevalent for cosmic strings. The final contribution is due to the damped acoustic oscillations of the intrinsic anisotropy and the Doppler effect around last scattering and it is this which we shall consider in this paper. This is not to say that the other contributions will have not effect over the range of angular scales considered, rather that there are other effects --- likely to be of similar amplitude --- which need to be taken into account. 

It is clear that in order to calculate the small-angle anisotropy, we must first calculate $\Theta_0$ and $V_{\rm b}$ around the time of last scattering and then deduce $\Theta_l$ using (\freestream). To do this we decouple the Boltzmann equation into its multipole moments,
\eqnam{\boltdecomp}
$$\eqalign{&\dot\Theta_0 = - \dot\Phi -{1\over 3}k\Theta_1\,,\quad
\dot\Theta_1 = k\big{(}\Theta_0+\Psi\big{)}-{2\over5}k\Theta_2-\dot\kappa (\Theta_1-V_{\rm b}
\big{)}\,, \cr &\dot\Theta_2 = {2\over 3}k\Theta_1-{3\over 7}k\Theta_3-{9\over 10}\dot\kappa\Theta_2\,,\quad
\dot\Theta_l =k\bigg{(}{l\over 2l-1}\Theta_{l-1}-{l+1\over 2l+3}
\Theta_{l+1}\bigg{)}-\dot\kappa\Theta_l\,,}\eqno(\new)$$
for $l>2$. As it stands (\boltdecomp) is not a complete system of equations, which can be rectified by introducing the Euler equation for the baryon velocity
\eqnam{\velocity}
$$\dot V_{\rm b}= -{\dot a\over a}V_{\rm b}+k\Psi+{\dot\kappa\over R}(\Theta_1-V_{\rm b})\,,\eqno(\new)$$
where $R=3\rho_{\rm b}/4\rho_{\gamma}$ is the baryon to photon ratio normalised to be $3/4$ at photon-baryon equality. Thomson scattering is highly efficient  at early times with $\dot\kappa$ large, therefore expanding (\boltdecomp) by ignoring ${\cal O}(k/\dot\kappa)$ yields the equations for tight-coupling regime. The effects of Silk damping can be investigated by expanding to higher order. If one ignores ${\cal O}(k^2/\dot\kappa^2)$ and also ${\cal O}(R^2)$, the equation for the intrinsic anisotropy is
\eqnam{\silkeom}
$${\ddot{\widehat\Theta_0}}+\left({\dot R\over 1+R}+ {8\over 27}k^2{1\over\dot\kappa}{1\over 1+R}
\right){\dot{\widehat\Theta_0}} + {1\over 3}k^2{1\over 1+R}{\widehat\Theta_0} = H(\eta)={1\over 3}k^2\left({\Phi\over 1+R}-\Psi\right)\eqno(\new)$$
and the baryon velocity is given by the dipole temperature anisotropy, $V_{\rm b}=\Theta_1=-3{\dot{\widehat\Theta_0}}/k$.

One can solve (\silkeom) for large $k$ using the WKB approximation and Green's method, 
\eqnam{\soln}
$$\eqalign{\left[1+R(\eta)\right]^{1/4}\widehat\Theta_0(\eta)=&e^{-k^2/k_{\rm s}^2(\eta,0)}\left[\widehat\Theta_0(0)\cos kr_{\rm s}(\eta)+{\sqrt{3}\over k}\left({\dot{\widehat\Theta_0}}(0)+{1\over 4}\dot R(0)\widehat\Theta_0(0)\right)\sin kr_{s}(\eta)\right]\cr
& +{\sqrt{3}\over k}\int_{0}^{\eta} d\eta^{\prime}\left[1+R(\eta^{\prime})\right]^{3/4}e^{-k^2/k_{\rm s}^2(\eta,\eta^{\prime})}\sin\left[kr_{\rm s}(\eta)-kr_{\rm s}(\eta^{\prime})\right] H(\eta^{\prime})\,,}\eqno(\new)$$
where $r_s(\eta)$ is the sound horizon distance and $k_{\rm s}^{-1}(\eta_2,\eta_1)$ is the Silk damping length, below which photon diffusion removes anisotropy,
\eqnam{\rsks}
$$r_{\rm s}(\eta)={1\over\sqrt{3}}\int_0^{\eta}{d\eta^{\prime}\over\sqrt{1+R(\eta^{\prime})}}\,,\quad k^{-2}_{\rm s}(\eta_2,\eta_1)={4\over 27}\int_{\eta_1}^{\eta_2}{d\eta^{\prime}\over\dot\kappa(\eta^{\prime})}{1\over 1+R(\eta^{\prime})}\,.\eqno(\new)$$
This solution has two parts: the transient whose amplitude and phase is dependent on the initial conditions  and the particular integral due to the forcing term. At the simplest level, the transient solution corresponds to the contribution from passive perturbations, while calculating the particular integral represents the effects of active perturbations by a convolution of the source term and the oscillatory terms which represent the acoustic waves with the appropriate damping coefficient\foot{\dag}{It should be noted that passive perturbations also create perturbations at horizon crossing and hence  the particular integral will also contribute. However, it can be seen that a source creating perturbations on lengthscales close to the horizon size will lead to a contribution close to being in phase with the initial conditions.}. The effect of Silk damping is clearly different for the passive and active perturbations contrary to the approach of refs.\refto{HSa,HSb}, since the damping length for active perturbations depends on when the perturbation is created.

In order to investigate the consequences of this effect on the microwave background, we shall make a number of simplifying assumptions. These assumptions allow simple calculations to be made, but the underlying physical principles which we want to illustrate will remain. First, we assume $R\approx 0$, which will be a good approximation in the low baryon content universes predicted by Big-Bang nucleosynthesis, since $R(\eta_*)\sim 30\Omega_{\rm b}h^2<<1$ \foot{\ddag}{This approximation ignores the effects of baryon drag\refto{HSd}, which lead to a power-law fall-off in the power spectrum at small angular scales. Since this effect is also due to the non-trivial time dependence of the gravitational potential it is likely to be of a similar size to the power-law fall-off discussed here, although the exact relative size will depend crucially on the baryon density.}. Also we assume only active perturbations, by ignoring the transient part of the solution.
Under these assumptions, the sound speed of the photon-baryon fluid is $1/\sqrt{3}$ and  
$$\eqalign{\widehat\Theta_0(\eta)=&{\sqrt{3}\over k}\int_0^{\eta}d\eta^{\prime}
e^{-k^2/k_{\rm s}^2(\eta,\eta^{\prime})}\sin\left[{k\over\sqrt{3}}(\eta-\eta^{\prime})\right]H(\eta^{\prime})\,,\cr {1\over\sqrt{3}}\Theta_1(\eta)=&-{\sqrt{3}\over k}\int_0^{\eta}d\eta^{\prime}
e^{-k^2/k_{\rm s}^2(\eta,\eta^{\prime})}\cos\left[{k\over\sqrt{3}}(\eta-\eta^{\prime})\right]H(\eta^{\prime})\,,}\eqno(\new)$$
where $H(\eta)=k^2(\Phi-\Psi)/3$ and $k^{-2}_{\rm s}(\eta_2,\eta_1)=(4/27)\int_{\eta_1}^{\eta_2}d\eta^{\prime\prime}/\dot\kappa(\eta^{\prime\prime})$.

At this stage, we must attempt to specify the nature of the source. This can be done using a structure function\refto{AlbSta}, a two-point correlation function of the source potentials, which determines the wavenumbers that are perturbed at a particular time. A simple example is $x^4F(x)\sim\delta(x-x_{\rm c})$, for which a single wavenumber $k=x_{\rm c}/\eta$ will be perturbed at conformal time $\eta$. More realistic examples of active perturbations will generate perturbations over a range of wavelengths. Of particular interest in this context is the asymptotic fall-off of the structure function: for small $x$ causality\refto{RW} suggests that $F(x)\sim {\rm const}$, whereas for large $x$ the fall-off is given by the characteristic shape of the anisotropies created by a particular model. Cosmic strings form line-like patterns\refto{KS} leading to $x^4F(x)\sim {\cal O}(x^{-2})$, whereas textures form spots implying that $x^4F(x)\sim{\cal O}(x^{-4})$. For definiteness, we use $F(x)=(x^2+a^2)^{-2}(1-bx+cx^2)^{-n}$ with $a=2.45$, $b=0.12$, $c=0.006$ and n=1 or 2. If $n=1$ the structure function may model a cosmic string network\refto{MACFb}, while $n=2$ will be described as a texture model, although in reality the precise values of $a$, $b$ and $c$ are likely to be somewhat different. These models have been chosen to exemplify the possible effects of the modified formalism, although they do have some relevance for the realistic models which one eventually wants to describe.

\figure{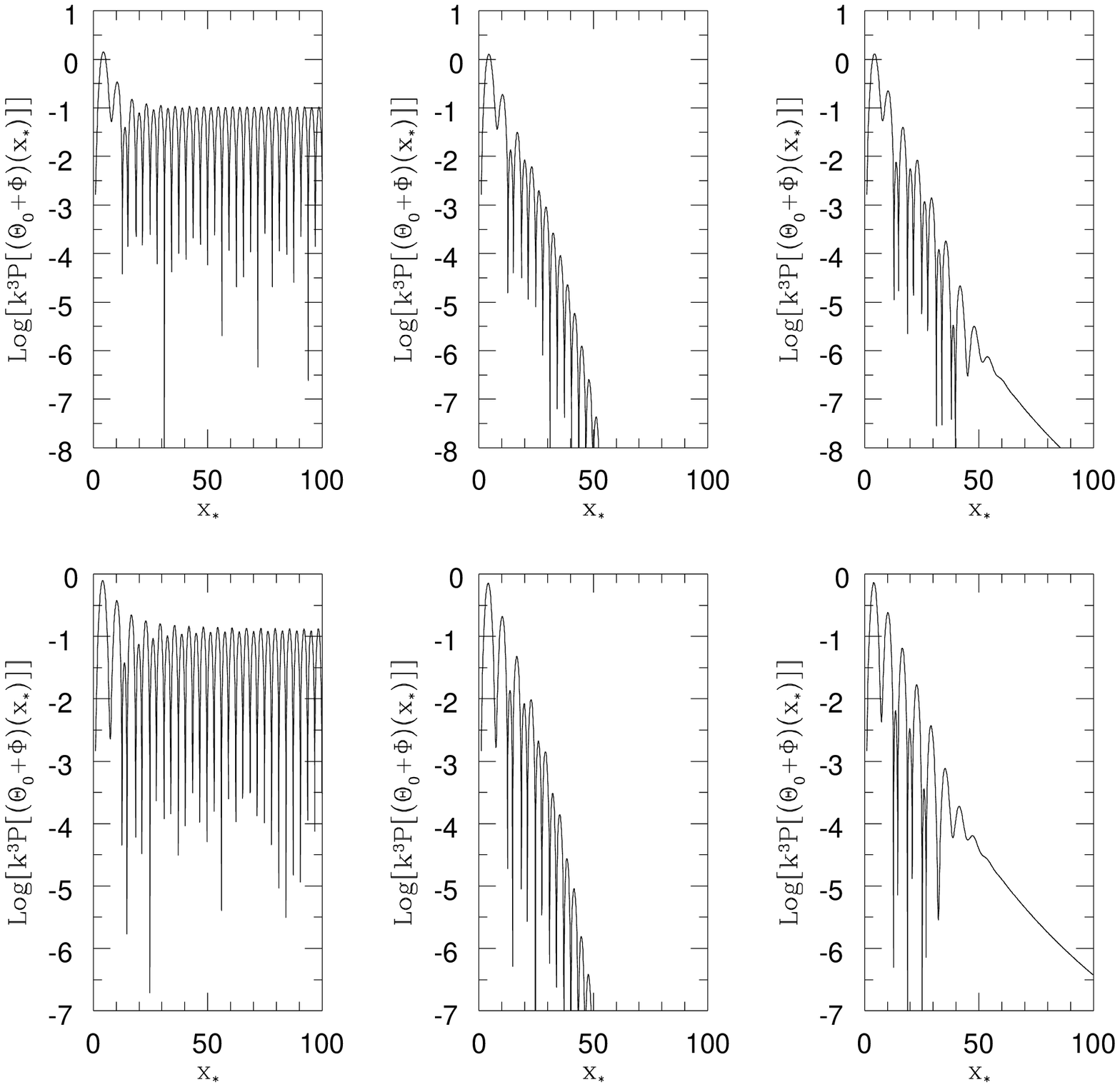}{3.1in}{3.1in}{0.90in}{1(a)}{The effects of the modified damping formalism compared to the tight coupling solution and the standard Silk damping formalism for coherent statistics. We plot the power spectrum of intrinsic component of the anisotropy on a log-linear scale. On the left, the tight coupling solution for the cosmic string (bottom) and texture structure (top) functions, in the centre the standard damping formalism and on the right the modified formalism. It is clear that the modified formalism allows much more power to remain on very small scales, with a power law fall-off.}

\figure{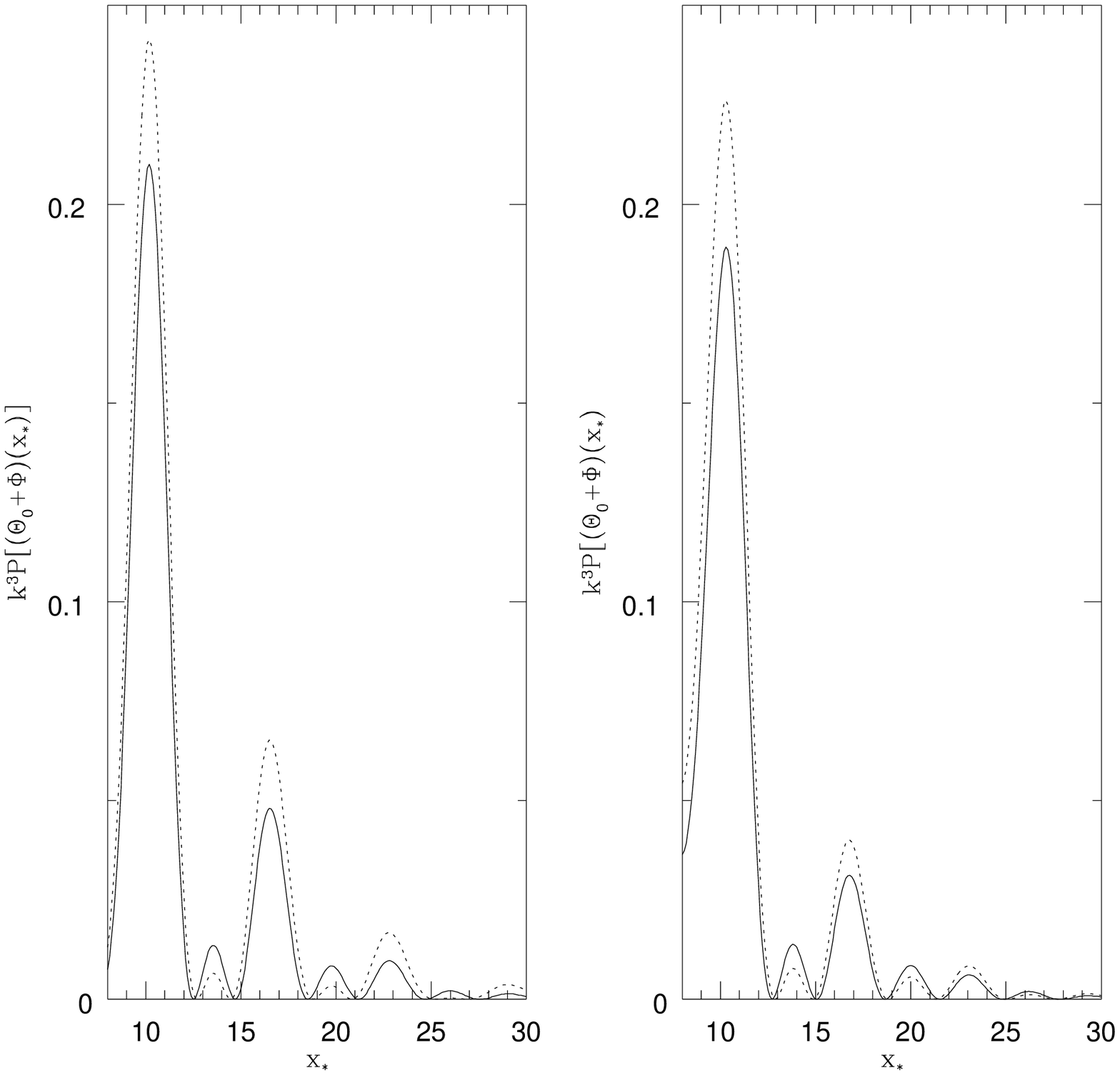}{3.1in}{3.1in}{0.90in}{1(b)}{The effects of the modified damping formalism (dotted line) compared to the standard Silk damping formalism (solid line) for coherent statistics. We plot the intrinsic anisotropy on a linear scale for, on the left the cosmic string model and on the right the texture model. This exhibits the differences in the peak height for the secondary peaks. Note that the scale starts at $x_*=8$, which ignores the first peak.}

\figure{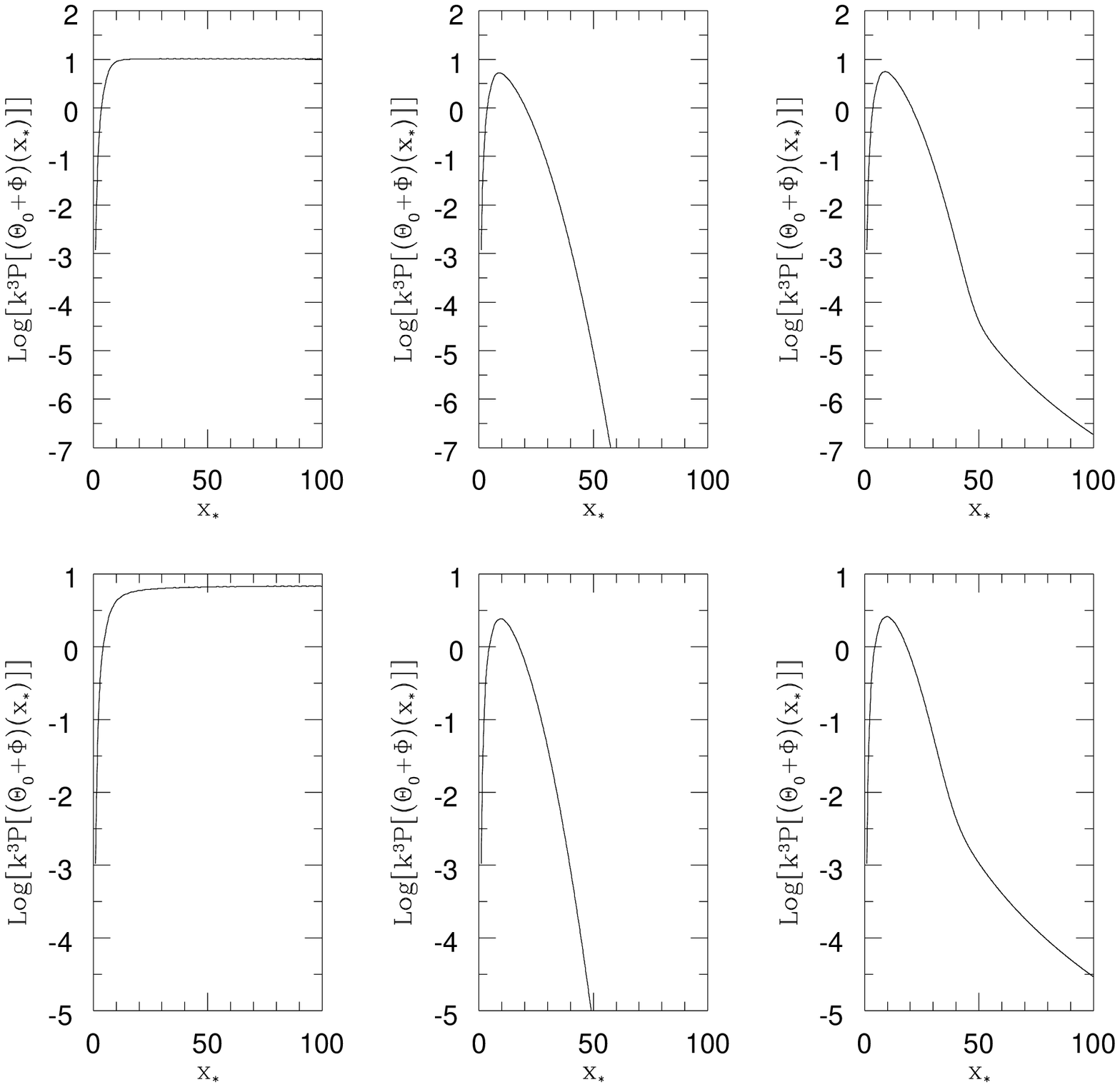}{3.1in}{3.1in}{0.90in}{2}{The intrinsic component of the anisotropy for incoherent statistics using the same conventions as fig.~1(a). Once again the modified damping formalism creates more power on small scales.}

\sectbegin{3}{Semi-analytic estimates}

\noindent For the purposes of this paper, we first concentrate on calculating the power spectrum (denoted by ${\cal P}$) of the intrinsic and Doppler contributions to the anisotropies, exhibiting the differences between the simple exponential suppression --- described as the standard Silk damping formalism --- and the more accurate modified damping formalism described in the previous section. This allows the study of the separate physical effects without confusion, within a calculationally more simple and intuitive framework.

In order to do this one must specify the type of statistics the source satisfies. A simple approximation often used is to assume that the source is totally coherent. In this case ${\cal P}(\Phi-\Psi)=\eta^3F(k\eta)$\refto{MACFa,MACFb} and hence the power spectrum of the intrinsic contribution to the anisotropy is given by
\eqnam{\cohpowermono}
$$k^{3}{\cal P}[\Theta_0+\Phi]={1\over 3}\left[\int_0^{x_{*}\sqrt{3}}
dx\,x^{3/2}[F(x)]^{1/2}\sin\left( x_*-{x\over\sqrt{3}}\right)\exp\left(-{3x_{*}^2\over \eta_*^2}k^{-2}_{\rm s}\left(\eta_*,{x\eta_*\over x_*\sqrt{3}}\right)\right)\right]^2\,,\eqno(\new)$$
as a function of the dimensionless parameter $x_*=k\eta_*/\sqrt{3}$. However, if the source is totally incoherent then various modifications to our understanding must be made\refto{MACFa}. In this case, 
\eqnam{\incoh}
$$k^{3}{\cal P}[\Theta_0+\Phi]={1\over 3}\int_0^{x_{*}\sqrt{3}}
dx\, x^4 F(x)\sin^2\left(x_*-{x\over\sqrt{3}}\right)\exp\left(-{6x_{*}^2\over \eta_*^2}k^{-2}_{\rm s}\left(\eta_*,{x\eta_*\over x_*\sqrt{3}}\right)\right)\,.\eqno(\new)$$
The difference between these two expressions is that the coherent approximation allows contributions to the anisotropy created at times very much before the last scattering to cancel out, whereas the incoherent approximation prevents any cancellations whatsoever, leading to the absence of secondary Doppler peaks\refto{MACFa,MACFb}. Neither of the approximations is likely to be a complete representation of the true physical situation on all scales, although one may be dominant. Here, we shall present results for both approximations, but the reader should note that the texture models are likely to be more coherent than string models. 

One can investigate the effects of damping by evaluating intrinsic anisotropy  power spectra (\cohpowermono) or (\incoh) for particular structure functions using a numerical integration routine and the simple model for the standard reionization history $(\Omega_0=1, \Omega_{\rm b}=0.05, h=0.5)$ presented in ref.\refto{HSa}. Figs.~1(a),(b) show the results for the coherent approximation. Both models have their first peak at $x_{*}\sim 4$ since that is only dependent on the coefficients $a$, $b$ and $c$ which have been kept constant. If damping is included, differences between the standard and modified spectra become noticeable for $x_{*}\gapp 10$ and dominate for $x_{*}\gapp 50$; the effect being very much stronger for strings since they create more power on the very smallest scales. As fig.~1(b) shows the peak heights are modified by an increasing amount as $x_*$ increases with, for example, in the cosmic string model the second, fourth and six peak heights being increased by 13\%, 25\% and 44\% respectively.  

The decay of the spectrum is approximately exponential for $x_*\lapp 50$ as in the passive case, but with a slightly different coefficient which leads to the modifications in the peak heights. However, for $x_*\gapp 50$ the behaviour of the spectrum is very different; power law decay with virtually no peaks. This distinction can be understood by realising that for $x_*\lapp 50$, a substantial proportion of the power is created before the last scattering epoch and hence the spectrum experiences almost the full effect of the damping, while for $x_*\gapp 50$ the power is created during the last few oscillations of the radiation fluid, preventing cancellations and hence peaks, with  the power law decay reflecting the large $x$ fall-off of the structure function. The same features are seen in the power spectrum for the texture model, but the modifications to the peak height are smaller and the transition to the power law regime is shifted to smaller scales $(x_{*}\sim 60)$ since less power is created on the smallest scales.

Fig.2 shows the equivalent spectra for the incoherent approximation. The tight coupling solutions are now totally devoid of any oscillations since no cancellations have taken place. This seems to lead to a further shift in the peak to about $x_*\sim 8$, due to perturbations created at earlier times feeding into smaller scales. However, the general picture of the modified formalism differing from the standard picture for $x_*\gapp 10$ remains.

\figure{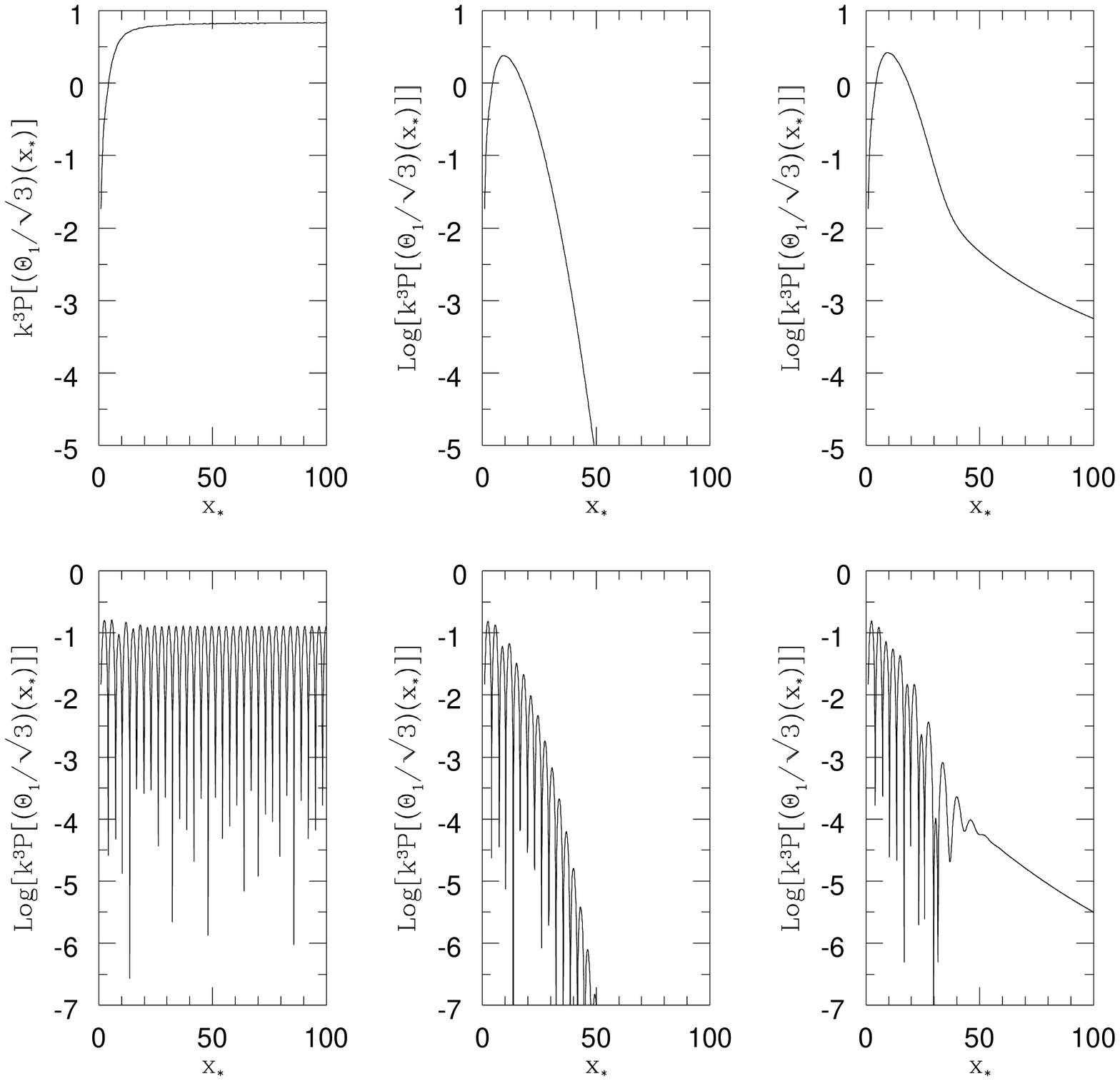}{3.1in}{3.1in}{0.90in}{3}{The effects of the new damping formalism on the Doppler contribution to the anisotropy. On the left, the tight coupling solution for a coherent (bottom) and incoherent (top) cosmic string structure function, in the centre the standard damping formalism and on the right the modified formalism. Notice that the difference between the standard and modified formalisms is more obvious than for the intrinsic contribution.}

The power spectrum of the Doppler (or dipole) component was also investigated by replacing the sine in (\cohpowermono) and (\incoh) with a cosine. We have already commented that the anisotropy on small angular scales is dominated by the last few oscillation of the radiation fluid. Formally this corresponds to around the region of the top limit of the integral in (\cohpowermono) and (\incoh). For the intrinsic component the integrand is zero at the upper limit, but the situation is very much different for the Doppler component which is totally out of phase with that of the intrinsic component. It will never be zero in the vicinity of the upper limit of the integral and hence the effect is enhanced. Fig.~3 illustrates this for the coherent and incoherent string models  Of course the Doppler component is suppressed with respect to the intrinsic anisotropy for $R>0$, but this will have significant implications for the spectrum of polarisation and density perturbations produced since they are created by the dipole anisotropy.

\figure{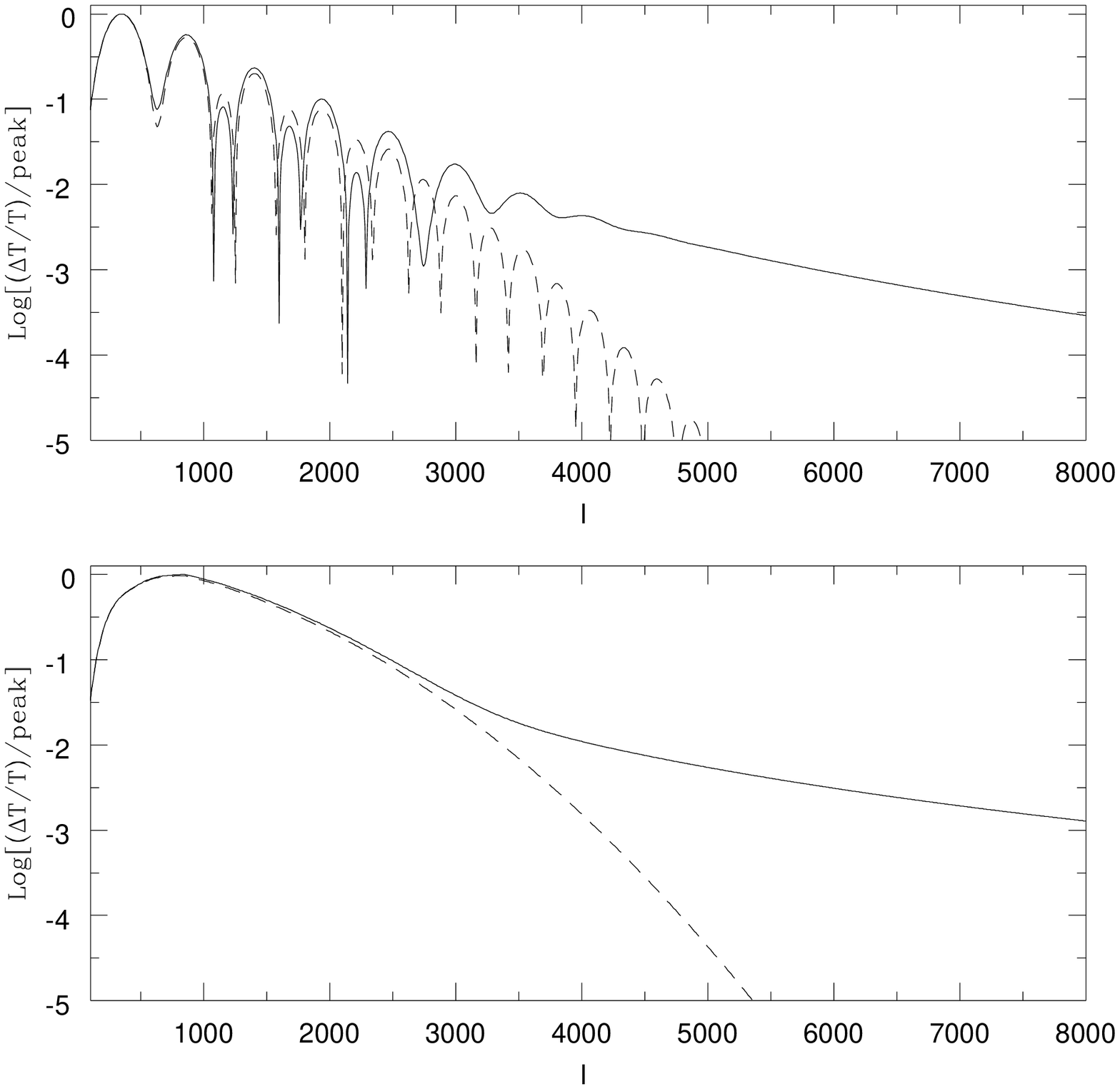}{3.1in}{3.1in}{0.90in}{4}{The angular power spectrum of the anisotropy due to the (dominant) intrinsic component. On the bottom, a sample incoherent cosmic string spectrum for the standard Silk damping approach (dotted line) and the modified damping formalism with a finite last scattering surface (unbroken line), which includes the effects of cancellation damping. The top illustrates the same spectra for the coherent string model. Notice the changes in the peak heights even on this log-linear scale, and the power-law rather than exponential tail.}

\sectbegin{4.}{Discussion and conclusions}

\noindent Until now we have concentrated on the power spectrum of the intrinsic and Doppler anisotropies. 
In order to calculate the power spectrum of the total anisotropy, we first assume that it is dominated by the intrinsic component, which is likely to be a good assumption for angular scales between $1^{\circ}$ and $1^{\prime}$---although note the discussion of the increased contribution from the dipole above. To deduce an expression similar to those provided in ref.\refto{HSa}, we must be careful to include the effects of cancellation on small scales due to finite width $\Delta\eta_*$ of the last scattering surface. This is represented by the rapidly oscillatory nature of the spherical Bessel function at large $k$. Averaging this effect gives $\ll j_l[k(\eta_0-\eta)]\rr\approx (k\Delta\eta_*)^{-1/2}j_l[k(\eta_0-\eta_*)]$ for $k\gapp (\Delta\eta_*)^{-1}$ and hence 
\eqnam{\ansi}
$${\Theta_l(\eta_0,k)\over 2l+1}={\sqrt{3}\over k}j_l[k(\eta_0-\eta_*)]\int_0^{\eta_0}d\eta^{\prime}H(\eta^{\prime})\sin
\left[{k\over\sqrt{3}}(\eta-\eta^{\prime})\right]{\cal D}(k,\eta^{\prime},\eta_0)\,,\eqno(\new)$$
where ${\cal D}(k,\eta^{\prime},\eta_0)$ is the equivalent of the acoustic visibility function in ref.\refto{HSa}, given by 
\eqnam{\acoustic}
$${\cal D}(k,\eta^{\prime},\eta_0)=\int_{\eta^{\prime}}^{\eta_0}d\eta
\dot\kappa e^{-\kappa(\eta,\eta_0)}\,e^{-k^2/k^2_{\rm s}(\eta,\eta^{\prime})}\,,\eqno(\new)$$ 
for $k<(\Delta\eta_*)^{-1}$
and multiplied by a factor of  $(k\Delta\eta_*)^{-1/2}$ for $k>(\Delta\eta_*)^{-1}$.

Fig.~4 shows a comparison between the standard damping mechanism and the modified damping mechanism for a finite width last scattering surface $(\eta_*/\Delta\eta_*\approx 6)$ including the effects of cancellation damping for the coherent and incoherent string models. We have assumed that the anisotropy is dominated by the intrinsic component and have used the simple relation $l=k\eta_0$ to calculate $(\Delta T/T)^2\propto \int dk\, k^2|\Theta_l(\eta_0,k)/(2l+1)|^2$. Although this calculation is approximate, it shows the important differences of the power spectra. It is clear that the level of anisotropy possible at very small angular scales is much larger with the modified damping formalism. 

To summarise, we have presented a simple, intuitive formalism to treat the effects of photon diffusion on the CMBR due to active sources.  
Simple models for cosmic string and texture scenarios illustrate that the effects of Silk damping mechanism are somewhat different to previously thought for scales smaller than the first acoustic peak, particularly in the case of strings. At this stage, we do not claim quantitative accuracy, rather that we have illustrated an effect which must be taken into account when making comparison between state of the art Boltzmann codes and semi-analytic methods similar to those described here, and that it should be included in  future work.

\nosectbegin{Acknowledgements}
 
\noindent Thanks to Andy Albrecht for impetus and encouragement. I have also benefited from conversations with Diego Harari, Joao Magueijo, Pedro Ferreira, Rob Caldwell, Neil Turok and Paul Shellard. I am supported by PPARC postdoctoral fellowship grant GR/K94799.



\def\hang{}


\def\jnl#1#2#3#4#5#6{\hang{#1 [#2], {\it #4\/} {\bf #5}, #6.}
									}


\def\jnlerr#1#2#3#4#5#6#7#8{\hang{#1 [#2], {\it #4\/} {\bf #5}, #6.
{Erratum:} {\it #4\/} {\bf #7}, #8.}
									}


\def\ibid#1#2#3#4#5#6#7#8{\hang{#1 [#2], {\it #4\/} {\bf #5}, #6; {\it ibid} {\bf #7}, #8.}
								}

\def\prep#1#2#3#4{\hang{#1 [#2], #4.}
									}


\def\book#1#2#3#4{\hang{#1 [#2], {\it #3\/} (#4).}
									}



\def\prl{Phys.\ Rev.\ Lett.}
\def\pr{Phys.\ Rev.}

\def\apj{Ap.\ J.}

\def\mn{M.$\,$N.$\,$R.$\,$A.$\,$S.}

\def\cupress{Cambridge University Press}

\def\cram{\vskip -2pt}

\nosectbegin{References}

\references

\baselineskip 10pt
\let\it=\nineit
\let\rm=\ninerm
\let\bf=\ninebf
\rm

\refis{ACFM}
\jnl{Albrecht A., Coulson D., Ferreira P. \& Magueijo J.}{1996}{}{\prl}{76}{1413}\cram

\refis{MACFa}
\jnl{Magueijo J., Albrecht A., Coulson D. \& Ferreira P.}{1996}{}{\prl}{76}{2617}\cram

\refis{MACFb}
\prep{Magueijo J., Albrecht A., Coulson D. \& Ferreira P.}{1996}{}{
astro-ph/9605047, {\it \pr} {\bf D} ~in press}\cram

\refis{kibb}
\jnl{Kibble T.W.B.}{1976}{}{J. Phys.}{A9}{1387}\cram

\refis{zeld}
\jnl{Zel'dovich Ya.B.}{1980}{}{\mn}{192}{663}\cram

\refis{vil}
\jnlerr{Vilenkin A.}{1981}{}{\prl}{46}{1169}{46}{1496}\cram

\refis{turok}
\jnl{Tuork N. \& Spergel D.N.}{1989}{}{\prl}{64}{2736}\cram

\refis{CT}
\jnl{Crittenden R.G. \& Turok N.}{1995}{}{\prl}{75}{2642}\cram

\refis{DGS}
\jnl{Durrer R., Gangui A. \& Sakellariadou M.}{1996}{}{\prl}{76}{579}\cram

\refis{silk}
\jnl{Silk J.}{1968}{}{\apj}{151}{459}\cram

\refis{HSa}
\jnl{Hu W. \& Sugiyama N.}{1995}{}{\apj}{444}{489}\cram

\refis{HSb}
\jnl{Hu W. \& Sugiyama N.}{1995}{}{\pr}{D51}{2599}\cram

\refis{HSc}
\jnl{Hu W. \& Sugiyama, N.}{1995}{}{\pr}{D50}{627}\cram

\refis{HSd}
\prep{Hu W. \& Sugiyama, N.}{1995}{}{astro-ph/9602019, {\it \apj} ~in press}\cram

\refis{SW}
\jnl{Sachs R.K. \& Wolfe A.M}{1967}{}{\apj}{147}{73}\cram

\refis{KS}
\jnl{Kaiser N. \& Stebbins A.}{1984}{}{Nature}{310}{391}\cram

\refis{ACSSV}
\prep{Allen B., Caldwell R., Shellard E.P.S., Stebbins A. \& Veeraraghavan S.}{1996}{}{{\it \prl} ~in press}\cram

\refis{bar}
\jnl{Bardeen J.M}{1980}{}{\pr}{D22}{1882}\cram

\refis{AlbSta}
\ibid{Albrecht A. \& Stebbins A.}{1992}{}{\prl}{68}{2121}{69}{2615}\cram

\refis{RW}
\jnl{Robinson J. \& Wandelt B.}{1996}{}{\pr}{D53}{618}\cram

\refis{VS}
\book{Vilenkin A. \& Shellard E.P.S.}{1994}{Cosmic strings and other topological defects}{\cupress}\cram

\refis{HK}
\jnl{Hindmarsh M.B. \& Kibble T.W.B.}{1995}{}{\it Rep. Prog. Phys.}{58}{477}\cram

\endreferences

\vfill
\end